\documentstyle[aps,pra,twocolumn]{revtex}

\begin{document}
\draft

\newcommand{\pp}[1]{\phantom{#1}}
\newcommand{\be}{\begin{eqnarray}}
\newcommand{\ee}{\end{eqnarray}}
\newcommand{\ve}{\varepsilon}
\newcommand{\vs}{\varsigma}
\newcommand{\Tr}{{\,\rm Tr\,}}
\newcommand{\pol}{\frac{1}{2}}
\newcommand{\ba}{\begin{array}}
\newcommand{\ea}{\end{array}}
\newcommand{\bear}{\begin{eqnarray}}
\newcommand{\eear}{\end{eqnarray}}
\title{
Thermostatistics based on Kolmogorov-Nagumo averages: 
Unifying framework for extensive and nonextensive generalizations
}
\author{Jan Naudts$^1$ and Marek Czachor$^{1,2}$}
\address{$^1$ Departement Natuurkunde, Universiteit Antwerpen UIA,
Universiteitsplein 1, B2610 Antwerpen, Belgium\\
$^2$ Katedra Fizyki Teoretycznej i Metod Matematycznych,
Politechnika Gda\'{n}ska, 80-952 Gda\'{n}sk, Poland\\
E-mail: Jan.Naudts@ua.ac.be and mczachor@pg.gda.pl}

\maketitle

\begin{abstract}
We show that extensive thermostatistics based on R\'enyi entropy and
Kolmogorov-Nagumo averages can be expressed in terms of Tsallis non-extensive
thermostatistics. We use this correspondence to generalize
thermostatistics to a large class of Kolmogorov-Nagumo means
and suitably adapted definitions of entropy. 
\end{abstract}


Generalized averages of the form 
\be
\langle x\rangle_\phi=
\phi^{-1}\Big(\sum_k p_k\phi(x_k)\Big),
\ee
where $\phi$ is an arbitrary continuous and strictly monotonic function,
were introduced into statistics by Kolmogorov
\cite{K} and Nagumo \cite{N}, and further generalized by de Finetti
\cite{dF}, Jessen \cite{J}, Kitagawa \cite{Kitagawa}, Acz\'el \cite{A} and
many others. Their first applications in information theory can be found
in the seminal papers by R\'enyi \cite{R1,R2} who employed them to
define a one-parameter family of measures of information
($\alpha$-entropies) 
\be
I_\alpha
&=&
\varphi_\alpha^{-1}\Big(\sum_k p_k\varphi_\alpha(\log_b\frac{1}{p_k})\Big)
=
\frac{1}{1-\alpha}\log_b\big(\sum_k p_k^\alpha\big).
\label{I-alpha}
\ee
The Kolmogorov-Nagumo (KN) function is here 
$\varphi_\alpha(x)=b^{(1-\alpha)x}$, a choice 
motivated by a theorem \cite{HL} stating that only affine or
exponential $\phi$ satisfy
\be
\langle x+C\rangle_\phi
=\langle x\rangle_\phi+C
\ee
where $C$ is a constant. Random variable 
\be
I_k=-\log_b p_k\label{hartley}
\ee
represents an
amount of information received by learning that an event of
probability $p_k$ took place \cite{S,W}; $b$ specifies units of information
($b=2$ corresponds to bits; below we use $b=e$
which is more common in the physics literature). 
$\alpha$-entropies were also derived in a purely pragmatic manner in 
\cite{R3} as measures of information for concrete
information-theoretic problems. 

R\'enyi's definition becomes more natural if one notices that 
KN-averages are invariant under $\phi(x)\mapsto A\phi(x)+B$ and one
replaces $\varphi_\alpha$ by 
\be
\phi_\alpha(x)=
\frac{e^{(1-\alpha)x}-1}{1-\alpha}
\equiv
\ln_\alpha[\exp(x)]\label{phi'}
\ee
where $\ln_\alpha(\cdot)$ is the deformed logarithm \cite{ln}
($\ln_1(\cdot)=\ln(\cdot)$). 

The above (original) derivation of $I_\alpha$ clearly shows the two
elements which led R\'enyi to the idea of $\alpha$-entropy: (1) 
one needs a generalized average and (2) the random variable one
averages is the logarithmic measure of information. The latter has a
well known heuristic explanation which goes back to Hartley \cite{H}:
To uniquely specify a single element of a set containing $N$ numbers
one needs $\log_2 N$ bits; but, if one splits the set into $n$ subsets
containing, respectively, $N_1,\dots,N_n$ elements ($\sum_iN_i=N$)
then in order to specify only in which set the element of interest is
located it is enough to have $\log_2 N-\log_2 N_i=\log_2 (N/N_i)$ bits of
information. The latter construction ignores the information encoded
in correlations between the subsets. For this reason typically one
needs less information if such correlations are present. The idea is
used in data-compression algorithms and is essential for the argument
we will present below. 

Although $\alpha$-entropies are occasionally used in statistical physics
\cite{Wehrl} it seems the same cannot be said of KN-averages. Thinking of the
original motivation behind generalized entropies one may wonder
whether this is not logically inconsistent. Constructing statistical
physics with $\alpha$-entropies one should consistently apply
KN-averaging to all random variables, internal energy included. 
Applying the procedure to thermostatistics one may expect to arrive
at a one-parameter family of equilibrium states which, in the limit
$\alpha\to 1$, reproduce Boltzmann-Gibbs statistics. 

During the past ten years it became quite clear that there is a need
for some generalization of standard thermostatistics, as exemplified
by the unquestionable success of Tsallis' $q$-thermodynamics
\cite{TC88}. Systems
with long-range correlations, memory effects or fractal boundaries
are well described by $q\neq 1$ Tsallis-type equilibria. Gradual
development of this theory allowed to understand that there is indeed
a link between generalized entropies and generalized averages. 
However, the averages one uses in Tsallis' statistics are the
standard linear ones but expressed in terms of the so-called escort
probabilities. So there is no direct link to KN-averages.

In what follows we present a thermostatistical theory based on
KN-averages. It deals with the problem
of maximizing average information under the constraint that
the average of some energy function has a given value. 
As we shall see there {\it is\/} a link between such a theory and
Tsallis' thermostatistics. Actually, many technical developments
obtained within the Tsallis scheme have a straightforward application
in the new framework. An important difference with respect to the Tsallis
theory is that we can obtain both non-extensive and extensive
generalizations so that one may expect the formalism will have still
wider scope of applications. 

We begin with the KN-average depending on parameters $p_k$ which
we shall later identify with escort probabilities.
$\alpha$-entropy defined with the help of the modified KN-function
(\ref{phi'}) is
\be
I_\alpha
&=&
\phi_\alpha^{-1}\Big(\sum_k p_k\phi_\alpha(I_k)\Big)
=
\phi_\alpha^{-1}\Big(\sum_k p_k\phi_\alpha(-\ln p_k)\Big)
\\
&=&
\phi_\alpha^{-1}\Big(\frac{\sum_k p_k^\alpha-1}{1-\alpha}\Big)
=
\phi_\alpha^{-1}\Big(\sum_k p_k \ln_\alpha(1/p_k)\Big).
\label{Renyi}
\ee
It is interesting that in the course of calculation of $I_\alpha$ the 
expression for the Dar\'oczy-Tsallis entropy \cite{Wehrl,TC88,DZ70} arises
\begin{equation}
S_\alpha(p)=\frac{\sum_k p_k^\alpha-1}{1-\alpha}.\label{tsallisalpha}
\end{equation}
This shows that in the context of KN-means there is an intrinsic
relation between $I_\alpha$ and $S_\alpha$:
\be
\phi_\alpha(I_\alpha)=S_\alpha. \label{a1}
\ee
Let us note that the formula 
\be
\phi_\alpha(I_k)=\ln_\alpha(1/p_k)\label{a2}
\ee
may hold also for other pairs $(\phi_\alpha,I_k)$,
with $\phi_\alpha$ not given by (\ref{phi'}),
and be valid even
for measures of information different from the Hartley-Shannon-Wiener
random variable $I_k=-\ln p_k$. 
The key assumption of the present paper is that the generalized
theory is characterized by the properties (\ref{a1}) and (\ref{a2}).
One can see (\ref{a2}) as a definition of $I_k$
in case $\phi_\alpha$ is given, or as a constraint on $\phi_\alpha$
if $I_k$ is given.

The generalized thermodynamics is obtained by maximizing $I_\alpha$
under the constraint of fixed internal energy
\be
\langle \beta_0 H\rangle_{\phi_\alpha}
=
\phi_\alpha^{-1}\Big(\sum_k p_k\phi_\alpha(\beta_0 E_k)\Big)
=
\beta_0 U
\ee
where $\beta_0$ is a constant needed to make the averaged energy
dimensionless. 
Equivalently, the problem may be reformulated as
maximizing (\ref{tsallisalpha}) under the constraint
\begin{equation}
\sum_kp_k \phi_\alpha(\beta_0E_k)=\phi_\alpha(\beta_0 U).
\label{constraint}
\end{equation}
This problem is of the type originally considered by
Tsallis \cite{TC88}. However, since then the formalism of
non-extensive thermostatistics has evolved. In particular, one has learned
\cite {TMP98} that the optimization problem should be
reparametrized using the so called {\it escort} probabilities.
The reason why one should do so is the following.
The standard thermodynamical relation for temperature $T$ is
\begin{equation}
\frac{1}{T}=\frac{{\rm d}S}{{\rm d}U}
\label{thermotemp}
\end{equation}
with $S$ and $U$, respectively, entropy and energy calculated using the
equilibrium averages. In generalized thermostatistics
this definition of temperature is not necessarily correct.
Recently has been shown \cite {AMPP01,TR01} that (\ref{thermotemp}) is
valid if the entropy is additive and must be modified in all other cases.
The reparametrization of non-extensive
thermostatistics, by introduction of escort probabilities, is
such that energy $U$ becomes generically an increasing function
of some (unphysical) temperature $T^*$
(see e.g.~Prop. 3.5 of \cite{NJ00}),
which is then related to physical temperature $T$.

The reparametrization is done by means of $q\leftrightarrow 1/q$ duality
\cite{TMP98,NJ01}.
Let
\begin{equation}
\rho_k=\frac{p_k^\alpha}{\sum_kp_k^\alpha}
\label{escort}
\end{equation}
Then one has clearly also the inverse relation
\begin{equation}
p_k=\frac{\rho_k^q}{\sum_k \rho_k^q}
\end{equation}
with $q=1/\alpha$.
The above optimization problem is now equivalent to maximizing
$S_q(\rho)$ under the constraint
\begin{equation}
\frac{\sum_k\rho_k^q \phi_{1/q}(\beta_0E_k)}{\sum_k \rho_k^q}
=\phi_{1/q}(\beta_0 U)
\end{equation}
This is so because $S_q(\rho)$ is maximal if and only if
$S_{1/q}(p)$ is maximal (see \cite{NJ01}).
The latter optimization problem is of the type studied in
the new style non-extensive thermostatistics \cite{TMP98}. 


We are now ready to solve the optimization problem.
The free energy $F$ is defined by
\begin{equation}
\beta_0F=\frac{\sum_k\rho_k^q \phi_{1/q}(\beta_0E_k)}{\sum_k \rho_k^q}
-\beta_0T^*S_q(\rho)
\end{equation}
Minima of $\beta_0F$, if they exist \cite{NJ00,NC01},
are realized for distributions of the form \cite {TMP98}
\begin{equation}
\rho_k\sim \frac{1}{[1+ax_k]^{1/(q-1)}}
\qquad \hbox{ if }1<q
\label{kappa}
\end{equation}
or
\begin{equation}
\rho_k\sim [1-ax_k]_+^{1/(1-q)}
\qquad \hbox{ if }0<q<1
\label{cutoff}
\end{equation}
Here $x_k=\phi_{1/q}(\beta_0E_k)$ and $[x]_+$ equals $x$ if $x$ is positive,
zero otherwise.
Expression (\ref{kappa}), with $1/(q-1)$ replaced by $1+\kappa$,
is called the kappa-distribution or generalized Lorentzian
distribution \cite{MZ00}.
There are several reasons why this distribution is of interest.
In the first place, the Gibbs distribution,
which determines the equilibrium average in the
standard setting of thermodynamics \cite{JE57},
is obtained in the limit $\kappa\rightarrow+\infty$,
or $q\rightarrow 1$.
The kappa-distribution is frequently used.
For example in plasma physics it is used to describe an excess
of highly energetic particles \cite{MVMH95}.
Typical for distribution (\ref{cutoff}) is that the probabilities $p_k$ 
are identically zero whenever $aE_k\ge 1$.
This cut-off for high values of $E_k$ is of interest in many areas of physics.
In astrophysics it has been used \cite{PP93} to describe stellar systems
with finite average mass.
A statistical description of an electron captured
in a Coulomb potential requires the cut-off to mask scattering
states \cite{LST95,NJ01}. In standard statistical mechanics the treatment of
vanishing probabilities requires infinite energies which lead to
ambiguities. These can be avoided if distributions of the type (\ref{cutoff})
are used.


The formulas that follow are based on results already found
in literature at many places, e.g.~in \cite{NJ00}.

Assume first that $\alpha=1/q$, $0<\alpha<1$. Then the equilibrium average
is the KN-average with $p_k$ given by
\begin{equation}
p_k=\frac{1}{Z_1}\frac{1}{[1+ax_k]^{1/(1-\alpha)}}.
\label{alphaless}
\end{equation}
$Z_1$ is given by
\begin{equation}
Z_1=\sum_k\frac{1}{[1+ax_k]^{1/(1-\alpha)}},
\end{equation}
and $x_k=\phi_\alpha(\beta_0 E_k)$.
The unknown parameter $a$ has to
be fixed in such a way that (\ref{constraint}) holds.
This condition can be written as
\begin{eqnarray}
\phi_\alpha(\beta_0 U)&=&\frac{1}{a}\left(\frac{Z_0}{Z_1}
-1\right)
\end{eqnarray}
with $Z_0$ given by
\begin{equation}
Z_0=\sum_k\frac{1}{[1+ax_k]^{\alpha/(1-\alpha)}}
\end{equation}
The entropy $I_\alpha$ follows from (\ref{tsallisalpha})
with (\ref{alphaless}). One obtains
\begin{equation}
\phi_\alpha(I_\alpha)=\frac{1}{1-\alpha}\left(\frac{Z_0}{Z_1^\alpha}-1\right).
\end{equation}
Temperature $T^*$ is given by (cf. Eq. (14) in \cite{NJ00})
\begin{equation}
\frac{1}{\beta_0T^*}=\frac{a\alpha}{1-\alpha}\,\frac{Z_1^2}
{Z_0^{(1+\alpha)/\alpha}}.
\label{tstarless}
\end{equation}
Now assume $\alpha>1$. Then the formulas become
\begin{equation}
p_k=\frac{1}{Z_1}[1-ax_k]_+^{1/(\alpha-1)}
\end{equation}
with $Z_1$ given by
\begin{equation}
Z_1=\sum_k[1-ax_k]_+^{1/(\alpha-1)}.
\end{equation}
The expressions for energy and entropy are
\begin{eqnarray}
\phi_\alpha(\beta_0 U)&=&\frac{1}{a}\left(1-\frac{Z_0}{Z_1}\right)
\end{eqnarray}
and
\begin{equation}
\phi_\alpha(I_\alpha)=\frac{1}{\alpha-1}\left(1-\frac{Z_0}{Z_1^\alpha}\right)
\end{equation}
with
\begin{equation}
Z_0=\sum_k[1-ax_k]_+^{\alpha/(\alpha-1)}
\end{equation}
Temperature $T^*$ is given by
\begin{equation}
\frac{1}{\beta_0T^*}=\frac{a\alpha}{\alpha-1}\,
\frac{Z_1^2}{Z_0^{(1+\alpha)/\alpha}}.
\end{equation}
Let us finally return to the specific case of R\'enyi's entropy,
i.e. $I_k$ and $\phi_\alpha$ given, respectively, by (\ref{hartley})
and (\ref{phi'}). This choice is particularly interesting since only
then the following three conditions are satisfied 
\be
\langle \beta_0 H+\beta_0E\rangle_{\phi_\alpha}
&=&
\langle \beta_0 H\rangle_{\phi_\alpha}+\beta_0 E\label{x1}\\
\langle \beta_0 H_{A+B}\rangle_{\phi_\alpha}
&=&
\langle \beta_0 H_A\rangle_{\phi_\alpha}
+
\langle \beta_0 H_B\rangle_{\phi_\alpha}\label{x2}
\\
I_\alpha(A+B)
&=&
I_\alpha(A)
+
I_\alpha (B)\label{x3}
\ee
where $A$ and $B$ are two uncorrelated noninteracting systems.
Condition (\ref{x1}) when combined with the explicit form of
equilibrium state means that equilibrium does not depend on the
origin of the energy scale. The remaining two conditions imply that
we have a one-parameter family of {\it extensive\/} generalizations
of the Boltzmann-Gibbs statistics, the latter being recovered in the
limit $\alpha\to 1$. For $\alpha=q^{-1}\neq 1$ we obtain the well known
Tsallis-type kappa-distributions but with energies $\beta E_k$ 
replaced by $\phi_\alpha(\beta_0 E_k)$. 

In general, the equilibrium probabilities are not of the product form
(there is one exception --- see below). 
The product form is of course also absent in the standard formalism
when there are correlations between subsystems. Nevertheless, if
the correlations are not too strong, then the system in equilibrium
is still extensive. This is expressed by stating that the so-called
thermodynamic limit exists. We expect that also in the present formalism
the thermodynamic limit exists, but this point has still to be studied.

Consider now the case $0<\alpha<1$ and $a=1-\alpha$.
This is a remarkable case because the equilibrium
distribution (\ref{alphaless}) becomes exponential.
Indeed, one verifies that
\begin{equation}
p_k=\frac{1}{Z_1}e^{-\beta_0 E_k}
\qquad\hbox{ with }
Z_1=\sum_ke^{-\beta_0 E_k}.
\end{equation}
Internal energy equals 
\begin{equation}
\beta_0U=\frac{1}{1-\alpha}\ln\left(
\frac{1}{Z_1}\sum_ke^{-\alpha\beta_0 E_k}
\right)
=I_\alpha-\ln Z_1.
\end{equation}
This means that for each system there exists a
particular temperature where the equilibrium state is factorizable.


Let us summarize the results. 

The formalism of thermostatistics based on 
KN-averages simultaneously generalizes Boltzmann-Gibbs
and Tsallis theories. As opposed to the Tsallis case, which is always
nonextensive, the KN-approach allows for a family of extensive 
generalizations. On the other hand, the family of extensive theories
leads to equilibrium states which share many properties with Tsallis
$q\neq 1$ distributions. Tsallis formalism enters the KN-formulation
also via the relation between $I_\alpha$ and $S_\alpha$. What is
surprising is that one should not simply identify $\alpha$ with $q$. The
correct relation $\alpha=1/q$ is suggested by the fact that the
probabilities $p_k$ are interpreted as escort probabilities. 
In the present paper the function $\phi_\alpha$ is kept constant
while the probabilities $p_k$ are varied. It could be interesting
to consider also the case where $\phi_\alpha$ is varied as well.

Of particular interest is the choice $\phi_\alpha(x)=\ln_\alpha(\exp x)$ 
since then the average information coincides with R\'enyi's entropy.
As proved by R\'enyi, his entropy, together with that of Shannon \cite {S},
are the only additive entropies. As shown in \cite {AMPP01,TR01}
additivity of entropy is a requirement for physical temperature $T$
to be defined by the usual thermodynamic relation (\ref{thermotemp}).
The formalism generalizes to other non-exponential choices of
$\phi$ provided the information measure is adapted in such a way
that (\ref{a1}) and (\ref{a2}) still hold. 
In this more general context entropy
is no longer additive and the definition of physical temperature $T$
by means of (\ref{thermotemp}) becomes problematic. A correct definition
could be derived along the lines of \cite {AMPP01} 
or \cite {TR01}. This problem requires further study.

In a natural way Tsallis' entropy appears as a tool for calculating
equilibrium averages. This offers the opportunity to
reuse the knowledge from Tsallis-like thermostatistics.
A tempting question is whether in each of the many applications
of Tsallis' thermostatistics one can find a natural KN-average
which maps the problem into the present formalism.

In an extended version of this Letter we shall discuss explicit
examples. Here we only mention that preliminary results for a
two-level system give satisfactory results. 
In particular, we checked that the $\alpha\to 1$ limit of equilibrium
distributions reproduces Boltzmann-Gibbs results,
and that the relation between $T$ and $T^*$ was found as in \cite{AMPP01}.
Of course, more complicated examples should be studied.
Examples like the one-dimensional Ising model could clarify
the issue of the thermodynamic limit. 

For the sake of completeness let us mention that R\'enyi's entropy has 
been studied already \cite{LMDS00} in relation with the escort probabilities
(\ref{escort}). One of the conclusions of that paper is that
they obtain the same results as in Tsallis' thermostatistics,
which is not a surprise since R\'enyi's entropy and Tsallis' entropy are
monotonic functions of each other. There is nor further relation with
the present work.


One of the authors (MC) wishes to thank the NATO for a research fellowship
enabling his stay at the Universiteit Antwerpen.


\end{document}